\documentclass[aps,prl,superscriptaddress,twocolumn]{revtex4}
\begin{document}

{\bf Reply to comment on ``Competing Interactions, the Renormalization Group and the Isotropic-Nematic Phase Transition''}

The focus of our work \cite{BaSt2007} was
to identify conditions for the presence of an isotropic-nematic phase
transition in the context of a generic
system with isotropic competing interactions.
 By taking into account non trivial angular momentum contributions from the  interaction, we found a second order isotropic-nematic phase transition at mean field level, which becomes a Kosterlitz-Thouless one \cite{KoTh1973} when fluctuations are taken into account. 

In his comment, Levin criticizes our results by showing that the low temperature fluctuations of a
stripe phase in 2d diverge linearly in the thermodynamic limit.  His
analysis is restricted to the stripe phase and, contrary to what is
suggested in the comment, does not apply to the central result of our letter which is
the existence of an isotropic-nematic phase transition. In fact, {\em
  as clearly anticipated by us in the letter} \cite{BaSt2007}, the
corresponding analysis of the fluctuations of the nematic order
parameter displays a logarithmic divergence leading to a low
temperature phase with quasi long range order.

In our model, despite of the involved calculations,  it is
straightforward to understand this fact. 
Introducing the nematic order parameter $\hat Q_{ij}=\alpha \left(\hat
n_i \hat n_j-\frac{1}{2}\delta_{i,j} \right)$ (where $\hat n_i=(\cos
\theta,
\sin \theta)$ is the director field) through a Hubbard-Stratonovich
transformation, it is possible to decouple the quartic $\phi$ terms.
Integrating out the $\phi$
field, we obtain the following long wavelength effective free energy
for the nematic order parameter: $F(\hat Q)=a_2/2\; Tr(\hat Q^2)+
a_4/4\; Tr(\hat Q^4)+ \rho/4\; Tr(\hat QD\hat Q)+\ldots$, where the
symmetric derivative tensor $D_{ij}=\nabla_i\nabla_j$ and $a_2$,
$a_4$ and $\rho$ are temperature dependent coefficients given in terms
of the parameters of the original model.  At mean field, the last term
is zero, and we find $\alpha=\sqrt{-a_2/a_4}$ for $a_2<0$, going
continuously to $\alpha=0$ for $a_2>0$. Note that any global
rotation of the order parameter costs no energy. Therefore,
parametrizing the order parameter by a modulus and an angle,  the
long wavelength angle fluctuations $\theta(x)$ dominate the low
energy  physics. Computing the free energy at lowest order in the
derivatives of the angle fluctuations we find  $\Delta F= \rho
\alpha^2 \int d^2x |\vec\nabla\theta|^2$, where $\Delta F$ is the
excess of free energy relative to the saddle point value. 
Therefore, the free energy of fluctuations corresponds to that of the  XY model. The only difference with the usual vector orientational order is that the system should have the symmetry 
$\theta\to \theta+\pi$ modifying the vorticity of the topological
defects.   Thus, one finds  for the angle fluctuations
$<\theta(x)\theta(x')>\sim \ln{[k_0(x-x')]}$, which in turn lead to an
algebraic decay of the order parameter correlations. In an extended
paper we will show the explicit dependence of the Frank constant
$K(T)=\rho \alpha^2$ with the parameters of our model $k_0$, $m$, $u_0$ and $u_2$. 
The conclusion is that the isotropic-nematic transition takes place by the Kosterlitz-Thouless mechanism of vortex (disclination) unbinding~\cite{KoTh1973}. This result agrees with the 
predictions of the well known KTHNY theory, that predicts the same kind of transition in a layered two-dimensional system~\cite{ToNe1981}. The main difference between our work and the KTHNY theory is that while KTHNY begin the analysis from an elastic energy valid at low temperatures already in the crystal phase, and explicitly add a term to take into account topological defects, we approach the transition form the disordered phase, allowing the possible emergence of spontaneous symmetry breaking. In this way, we make contact with more microscopic parameters, clarifying in some way the role of competing interactions in the development of orientational phases, contrary to what is suggested in the comment. It is well known that Ginzburg-Landau functionals similar to the one explored by us can be obtained from
more microscopic interactions like those in ultrathin magnetic films
\cite{GaDo1982} and copolymers in 2d~\cite{Haetal2002}. 

In conclusion, we have shown that the model introduced in
 equations (1) and (7) of Ref. \cite{BaSt2007}  undergoes an
 isotropic-nematic phase transition in the Kosterlitz-Thouless 
universality class.

We acknowledge CNPq and FAPERJ for partial financial support.

Daniel G.\ Barci\\
Departamento de F{\'\i}sica Te\'orica,
Universidade do Estado do Rio de Janeiro, Rua S\~ao Francisco Xavier 524, 20550-013,  Rio de Janeiro, RJ, Brazil.

Daniel A.\ Stariolo\\
Departamento de F{\'\i}sica, Universidade Federal do Rio Grande do Sul, CP 15051, 91501-970, Porto Alegre, RS, Brazil

\end{document}